\documentclass[letterpaper]{IEEEtran}
\input{preamble}
\title{Duality between Feature Selection\\ and Data Clustering}
\author{Chung Chan, Ali Al-Bashabsheh, Qiaoqiao Zhou and Tie Liu
	\thanks{Preliminary work has been submitted to \cite{chan16allerton}.}
	\thanks{C.\ Chan (email: cchan@inc.cuhk.edu.hk, chungc@alum.mit.edu),
		A.\ Al-Bashabsheh and Q.\ Zhou are with the Institute of Network Coding at the
		Chinese University of Hong Kong, the Shenzhen Key Laboratory of
		Network Coding Key Technology and Application, China, and the
		Shenzhen Research Institute of the Chinese University of Hong
		Kong.
	}
	\thanks{T.\ Liu is with the Department of Electrical and Computer
		Engineering, Texas
		A\&M University, College Station, TX 77843 USA (email: tieliu@tamu.edu).}
	\thanks{The work is supported in part by a grant from University Grants Committee of
		the Hong Kong Special Administrative Region, China (Project
		No. AoE/E-02/08), Shenzhen Research Fund (KQCX20130628164008004) and Shenzhen Key Laboratory of Network Coding
		Key Technology and Application, Shenzhen, China
		(ZSDY20120619151314964).}
	\thanks{The work of T. Liu was supported in part by the National
		Science Foundation under Grant CCF-13-20237. Part of the work was
		done while T. Liu was visiting the Institute of Network Coding at
		the Chinese University of Hong Kong.}
	\thanks{The work of C.\ Chan was supported in part by the University
		Grants Committee of the Hong Kong Special Administrative Region,
		China (Project No. 14200714).}}
\begin{document}
	
\IEEEoverridecommandlockouts
\maketitle

\begin{abstract}
	The feature-selection problem is formulated from an
	information-theoretic perspective. We show that the problem
	can be efficiently solved by an extension of the recently proposed
	info-clustering paradigm. This reveals the fundamental
	duality between feature selection and data clustering,
	which is a consequence of the more general duality between
	the principal partition and the principal lattice of
	partitions in combinatorial optimization.
\end{abstract}

\section{Introduction}

Many problems in machine learning are, in essence, the devising of a parametrized model that provides a good
approximation to the functional dependency between a set of input variables (features)
and an output (dependent) variable.%
\footnote{Depending on the context, the input variables are some times referred to as features in
	the machine learning literature and independent variables or regressors in regression analysis. In this
	work, we will refrain from the use of the term ``independent variables'' in the context of
	regression and reserve the term to refer to statistical independence between a set of random
	variables.}
The model parameters are often determined/estimated using a training set of points, where each
point is a pair consisting of a sample (i.e., a configuration) of the input variables and the
corresponding output value.
The set of features often contains irrelevant features to the output variable, which results in a
high processing complexity and overfitting (due to the limited size of the training set).
The feature selection problem is an attempt to resolve the above issues by selecting
the features that are most relevant to the output variable. 
This of course raises the two questions of what is meant by ``relevant'' and how can one determine such
relevant features. Shannon's mutual information~\cite{shannon48} was considered in \cite{battiti94} for the feature selection problem. It was also recognized that such a natural formulation~\cite[FR$n$--$k$]{battiti94} is impractical without further relaxation, owing to the high computational and sample complexity in estimating the mutual information for a large set of features from data. Hence, subsequent information-theoretic approaches such as \cite{bennasar15} have been focusing on finding good heuristics to solve the problem approximately. 

Another prominent problem in machine learning is the clustering problem. In a broad sense, this is
the problem of dividing a set of objects into groups such that elements in the same group are
similar/relevant to each other and elements from different groups are dissimilar/irrelevant to each
other.
Given a mathematically justifiable notion of similarity/relevance for clustering (see \cite{chan16cluster} for details),
then one may, at least intuitively,
provide a satisfying answer to the two questions above. 
Namely, one can treat the features and the dependent variable as the objects in hand, identify the
cluster that contains the dependent variable, and declare the remaining elements in the same cluster as
the most relevant features.
While in general this remains an intuition that may lack mathematical rigor,
in the special case when the features are statistically independent, 
we prove a duality theorem between the feature selection and data clustering problems
that will provide a precise mathematical explanation of the intuition above. 


The underlying pinnings to the feature selection and data clustering duality in this work are two
mathematical structures called the principal partition (PP) (see, e.g., \cite{fujishige-pp-revisited} for
an overview of related works to the PP)  and the principal lattice of partitions (PLP) \cite{narayanan97} of a submodular function.
%
%
%
Both the PP and PLP are polynomial-time computable, as will be pointed out in place.
%
The recognition of a link between the PLP (more precisely, a subset of the PLP) and the clustering
problem was made in \cite{nagano10}, which led to an efficient algorithm that provides a partial
solution to the hard $k$-clustering problem. The detailed connection was discussed in \cite{chan16cluster}.
%
In \cite{nagano11} the PP (more precisely, a subset of the PP) was linked to the size-constrained
submodular function minimization problem, which led to an efficient algorithm that provides a partial
solution to the problem. 

In this work, we connect the (entire) PP to the feature selection problem
(by showing that an element of the PP is a solution to the feature selection problem)
and connect the (entire) PLP to the data clustering problem
(by showing that an element of the PLP is a solution to the clustering problem).
When the features are independent, we prove a one-to-one correspondence between the PP and PLP,
thereby a duality between the feature selection and the clustering problems. (More precisely, the
duality is between the solutions of the two problems that are captured by the PP and PLP.) 

We remark that the duality result can be extended to more general submodular function. The current duality can be viewed as the special case when the entropy function is taken to be the submodular function and the modularity is the statistical independence among the features.  The only other duality we are aware of 
is in \cite{patkar-narayanan:92}, which gave the fastest algorithm at the time for the computation
of the PP of a graph. (By first computing the PLP of the graph and then constructing the PP via this duality.) However, that result cannot be put in the same category as the current result because it considers the PLP and PP for different submodular functions of a graph, namely, the graph cut function evaluated over subsets of vertices and the rank function of the cycle matroid of the graph evaluated over subsets of edges instead of the vertices. The duality result appears to exploit the graphical structure; there seems to be no natural extension of such result beyond graphs.


\section{Motivation}

As a motivation for the duality result, we will consider a simple example involving
two independent random variables $\RX_1$ and $\RX_2$, and a third random variable $\RY$.
For the feature selection problem, let $\RX_1$ and $\RX_2$ be the features
and $\RY$ be the dependent variable. One is
interested in selecting subsets of the features that are highly correlated with the
dependent variable.
More precisely, feature $i\in \Set {1,2}$ is the best feature if it maximizes
Shannon's mutual information~\cite{yeung08}:
\begin{align*}
	\max_{i\in \Set{1,2}} I(\RY \wedge \RX_i).
\end{align*}
As an illustration, assume the random variables are such that
\begin{subequations}
	\label{eq:eg:regress}
\begin{align}
		\RY=(\RX_1,\RX_2) \kern1em \text{with } &I(\RX_1\wedge \RX_2)=0 \text{ and}\label{eq:eg:regress:1}\\ &H(\RX_1)=2>H(\RX_2)=1. \label{eq:eg:regress:2}
\end{align}
\end{subequations}
The first variable $\RX_1$ is a better feature than $\RX_2$ as it shares more mutual information with the dependent variable $\RY$.

For the data clustering problem, we consider the info-clustering paradigm in \cite{chan16cluster}
which clusters a set of random variables according to their multivariate mutual information (MMI).
As an example, let $\RZ_0$, $\RZ_1$ and $\RZ_2$ be the three random variables we want to
cluster. Given a threshold $`g\in `R$, a cluster  is a subset
\begin{align*}
	B \subseteq \Set{0,1,2} : \abs {B}>1, I(\RZ_B)>`g, \forall B'\supsetneq B, I(\RZ_{B'})\leq `g,
\end{align*}
where $I(\RZ_B)$ is the multivariate mutual information (MMI) defined in~\cite{chan15mi} (to be introduced in \eqref{eq:MMI}).
In other words, a cluster is an inclusion-wise maximal subset of consisting of at
least two random variables with strictly more than $`g$ amount of mutual information.
In the above, the MMI measures the
mutual information among multiple random variables and may be viewed as an extension of Shannon's
mutual information from the bivariate to the multivariate case.

For simplicity, consider the example
\begin{align}
	\RZ_0=\RY, \RZ_1=\RX_1 \text{ and } \RZ_2=\RX_2,\label{eq:eg:cluster}
\end{align}
with $\RY$, $\RX_1$ and $\RX_2$ satisfying \eqref{eq:eg:regress}. Then, for $B\subseteq \{0,1,2\}$
with $|B|\geq 2$, the MMI can be calculated to be 
\begin{align}
	I(\RZ_B)=\begin{cases}
		0 & B=\Set {1,2}\\
		H(\RZ_1) & B=\Set {0,1}\\
		H(\RZ_2) & B\in \Set {\Set {0,1,2}, \Set {0,2}}.
	\end{cases}
	\label{eq:eg:I}
\end{align}
For instance, $I(\RZ_{\Set {0,2}})=H(\RX_2)$ because $\RX_2$ is the information shared among all
$\RZ_i$'s. 
In particular, $I(\RZ_{\Set {1,2}})=0$ because $\RZ_1=\RX_1$ and $\RZ_2=\RX_2$ are independent. $I(\RZ_{\Set {0,1}})=I(\RZ_0\wedge \RZ_1)=H(\RX_1)$ because $\RX_1$ is the
information shared between $\RZ_0$ and $\RZ_1$. Similarly, $I(\RZ_{\Set {0,1,2}})=I(\RZ_0\wedge \RZ_2)=H(\RX_2)$. The fact that $I(\RZ_{\Set{0,1,2}})=H(\RX_2)$, however, requires a more detailed understanding of the MMI. A concrete operational meaning~\cite{chan15mi} is through the secret key agreement problem, that $I(\RZ_{\Set{0,1,2}})$ is the maximum rate of secret key that can be agreed upon mutually among three users who observe privately the discrete memoryless sources $\RZ_0$, $\RZ_1$ and $\RZ_2$ respectively. An alternative mathematically appealing interpretation is the residual independence relation in \cite[Theorem~5.1]{chan15mi}: $`g=H(\RX_2)$ satisfies
\begin{align*}
	`1[H(\RZ_{\Set{0,1,2}})-`g`2] = \sum_{i=0}^{2}`1[H(\RZ_{i})-`g`2],
\end{align*} 
which is called the RIR because the total randomness on the L.H.S.\ after removing $`g$ is equal to the sum of the individual randomness of each random variable on the R.H.S.\ after removing $`g$. The equality can be taken to mean that there is no overlapping (mutual information) left in the residual randomness after removing $`g$, and so $`g$ reflects the amount of information mutual to the three random variables. A figure illustrating this can be found in \cite[Section~III--A]{chan16cluster}, which can be viewed as a natural extension of the well-known Venn-diagram interpretation of Shannon's mutual information~\cite{cover2012elements}.

Based on \eqref{eq:eg:I}, for $`g< H(\RX_2)$, the entire set $\Set
{0,1,2}$ of random variables is a cluster because it is trivially maximal and it satisfies the required threshold constraint by
\eqref{eq:eg:regress}, i.e.,
\begin{align*}
	I(\RZ_{\Set{0,1,2}})= H(\RX_2) > `g.
\end{align*}
By the same reasoning, when $H(\RX_2) \leq \gamma < H(\RX_1)$, the set $\Set {0,1}$ is a cluster. Note that even though
$\Set{0,2}$ satisfies the threshold constraint for $`g<H(\RX_2)$, it is not considered as a cluster because it is not maximal. More importantly, if the set $\Set{0,2}$ were a cluster, then it would be inconsistent with the cluster $\Set{0,1}$ which can be taken to assert that $\RZ_0$ shares more information with $\RZ_1$ (the element in the same cluster) than with $\RZ_2$ (the element outside the cluster).

The duality between feature selection and data clustering is simply that: \emph{as the threshold $`g$
increases, the dependent variable clusters with a smaller set of more relevant features.} In the
current example, with $`g$ large enough, i.e., exceeding $H(\RX_2)$, the better feature $\RX_1$ is
identified by the cluster $\Set {0,1}$, which groups the dependent variable $\RZ_0=\RY$ with the
feature $\RZ_1=\RX_1$. In this work, we extend the duality result to the case
allowing any number of independent random variables (as features) and any correlation between the dependent variable and
features.

\section{Info-clustering formulation}


\begin{figure*}
	\centering
	\subcaptionbox{$\hat{h}_{`g}(V)$ vs $`g$ in \eqref{eq:hat h} for \eqref{eq:eg:cluster}.\label{fig:eg:h}}{
		{\def\u{4}
			\tikzstyle{point}=[draw,circle,minimum size=.2em,inner sep=0, outer sep=.2em]
			\begin{tikzpicture}[x=.8em,y=.8em,>=latex]
			\draw[->] (0,-0.2*\u) -- (0,4*\u) node [label=right:$\hat{h}_{`g}(V)$] {};
			\draw[->] (0,0) -- (3.5*\u,0) node [label=right:$`g$] {};
			\foreach \i/\ya/\xa/\yb/\xb/\lp/\lb in {
				1/3.5/0.833/-0.5/2.166/left/{}, 
				2/3.5/0.25/-0.5/2.25/left/{}, 
				3/3/0/-0.5/3.5/left/{}, 
				4/3.5/0.75/-0.5/2.75/left/{$\begin{aligned}h_{`g}[\Set{\Set{0,2},\Set{1}}]&=\\&\kern-4em H(\RX_{\Set{1,2}}) +H(\RX_1)\kern.3em-2`g\kern-.1em\end{aligned}$},
				5/3.5/1.25/-0.5/3.25/right/{$\begin{aligned}h_{`g}[\Set{\Set{0},\Set{1,2}}]&=\\ &\kern-4em 2H(\RX_{\Set{1,2}})-2`g\end{aligned}$}}
			\draw[dashed,gray] (\xa*\u,\ya*\u)  node [inner sep=0,outer sep=0,label={[label distance=0em]\lp:{\scriptsize\lb}}] {} -- (\xb*\u,\yb*\u);
			\path (1*\u,2*\u) node (1) [point,red,thick,label=above right:{}] {};
			\path (2*\u,0*\u) node (2) [point,red,thick,label=above right:{}] {};
			\draw[dotted] (2*\u,0*\u) to (2*\u,-0.2*\u) node [label={[shift={(-1em,-1em)}]{\scriptsize$H(\RX_1)$}}] {};
			\draw[dotted] (1) to (1*\u,-0.2*\u) node [label={[shift={(-1em,-1em)}]{\scriptsize$H(\RX_2)$}}] {};
			\draw[-,thick,blue] (0,3*\u) to node [below left] {\scriptsize $h_{`g}[\Set{\Set {0,1,2}}]=H(\RX_{\Set{1,2}})-`g$} (1) to node [below left] {\scriptsize $h_{`g}[\Set{\Set{0,1},\Set{2}}]=H(\RX_{\Set{1,2}})+H(\RX_2)-2`g$} (2) to node [below left] {\scriptsize $h_{`g}[\Set{\Set{0},\Set{1},\Set{2}}]=2H(\RX_{\Set{1,2}})-3`g$} (2.166*\u,-0.5*\u);
			\end{tikzpicture}}}
	\subcaptionbox{$f^*(`g)$ vs $`g$ in \eqref{eq:regress} for \eqref{eq:eg:regress}.\label{fig:eg:f}}{
		{\def\u{4}
			\tikzstyle{point}=[draw,circle,minimum size=.2em,inner sep=0, outer sep=.2em]
			\begin{tikzpicture}[x=.8em,y=.8em,>=latex]
			\draw[->] (0,0*\u) -- (0,3.5*\u) node [label=right:$f^*(`g)$] {};
			\draw[->] (0,0) -- (3*\u,0) node [label=right:$`g$] {};
			\foreach \i/\ya/\xa/\yb/\xb/\lp/\lb in {
				1/0/0/0/3/left/{}, 
				2/1/0/0/1/left/{$f_{`g}(\Set{2})=H(\RX_2)-`g$}, 
				3/2/0/0/2/left/{}, 
				4/3/0/0/1.5/left/{}}
			\draw[dashed,gray] (\xa*\u,\ya*\u)  node [inner sep=0,outer sep=0,label={[label distance=0em]\lp:{\scriptsize\lb}}] {} -- (\xb*\u,\yb*\u);
			\path (1*\u,1*\u) node (1) [point,red,thick,label=above right:{}] {};
			\path (2*\u,0*\u) node (2) [point,red,thick,label=above right:{}] {};
			\draw[dotted] (2*\u,0*\u) to (2*\u,-0.2*\u) node [label={[shift={(-1em,-1em)}]{\scriptsize$H(\RX_1)$}}] {};
			\draw[dotted] (1) to (1*\u,-0.2*\u) node [label={[shift={(-1em,-1em)}]{\scriptsize$H(\RX_2)$}}] {};
			\draw[-,thick,blue] (0,3*\u) to node [above right] {\scriptsize $f_{`g}(\Set{1,2})=H(\RX_{\Set{1,2}})-2`g$} (1) to node [above right] {\scriptsize $f_{`g}(\Set{1})=H(\RX_1)-`g$} (2) to node [below right] {\scriptsize $f_{`g}(`0)=0$} (3*\u,0);
			\end{tikzpicture}}}
	\caption{Plots of \eqref{eq:hat h} and \eqref{eq:regress} for the example \eqref{eq:eg:regress} and \eqref{eq:eg:cluster} under the mapping \eqref{eq:map}.}
	\label{fig:eg}
\end{figure*}

In this section, we first introduce the general info-clustering formulation in \cite{chan16cluster} and then extend it slightly for the desired duality result. The framework considers any number of random variables with any joint distribution. More precisely, let $\RZ_V:=(\RZ_i\mid i\in V)$ be a finite vector of random variables to be clustered, and let $\Pi(V)$ be the collection of partitions of $V$ into non-empty disjoint sets. The set of clusters at a real-valued threshold $`g\in `R$ is defined in \cite[Definition~2.1]{chan16cluster} as
\begin{subequations}
	\label{eq:cluster}
\begin{align}
	\pzC_{`g}(\RZ_V) &:=\{B\subseteq V\mid \abs {B}>1,I(\RZ_B)>`g,\label{eq:cluster:1}\\
	&\kern1em \not\exists B'\supsetneq B, I(\RZ_{B'})>`g\label{eq:cluster:2}
		\},
\end{align}
\end{subequations}
where $I(\RZ_B)$ is the MMI defined as~\cite{chan15mi}
\begin{align}
	I(\RZ_B):=\min_{\substack{\mcP \in \Pi(B):\\ \abs {\mcP}>1}}
	\frac 1{\abs {\mcP}-1} \underbrace{D`1(\extendvert{P_{\RZ_V}\|\prod_{C\in \mcP} P_{\RZ_C}}`2)}_{=\sum_{C\in\mcP} H(\RZ_C)-H(\RZ_B)}.\label{eq:MMI}
\end{align}
In the bivariate case when $V=\Set{1,2}$, the MMI reduces to Shannon's mutual information $I(\RZ_1\wedge \RZ_2)$ with $\mcP=\Set{\Set {1},\Set {2}}$. \emph{The MMI naturally extends Shannon's mutual information to the multivariate case}, with concrete operational meanings in secret key agreement and undirected network coding~\cite{chan10md,chan11isit,chan15mi}. In the above, \emph{\eqref{eq:cluster:1} is the threshold constraint that requires the random variables in a cluster to share at least $`g$ amount of information, while the non-existence condition in \eqref{eq:cluster:2} requires the cluster to be inclusion-wise maximal.}

It was shown in~\cite{chan16cluster} that the clustering solution of \eqref{eq:cluster} is given by a mathematical structure called the principal lattice of partitions (PLP) introduced by \cite{narayanan90}. 
More precisely, we say that a set function $h:2^V\to `R$ is submodular~\cite{schrijver02} if for all $B_1,B_2\subseteq V$,
\begin{align}
	h(B_1)+h(B_2)\geq h(B_1\cup B_2)+h(B_1\cap B_2).\label{eq:submodular}
\end{align}
The function $h$ is said to be supermodular if the inequality above is reversed, and  modular if
equality holds. It follows that $-h$ is supermodular iff $h$ is submodular, while $h$ is modular iff
it is both submodular and supermodular. The entropy function 
\begin{align}
	h(B):=H(\RZ_B) \kern1em \text{for } B\subseteq V,\label{eq:h}
\end{align}
for instance, is known to be submodular~\cite{fujishige78}, and so is the residual entropy function~\cite{chan15mi} 
\begin{align}
	h_{`g}(B):=h(B)-`g.\label{eq:hg}
\end{align} 
More generally, a constant function is modular and the sum of submodular functions is submodular. For the submodular function $h_{`g}$, the Dilworth truncation~\cite{schrijver02} (evaluated at $V$) is
\begin{align}
	\hat h_{`g}(V):=\min_{\mcP\in \Pi(V)} \underbrace{\sum_{C\in \mcP} \overbrace{[H(\RZ_C)-`g]}^{h_{`g}(C)}}_{h_{`g}[\mcP]:=}, \label{eq:hat h}
\end{align}
the optimal partitions to which for different values of $`g\in `R$ is called the PLP~\cite{narayanan90}. As an illustration, \figref{fig:eg:h} shows a plot of $\hat h_{`g}(V)$ against $`g$ for the example in~\eqref{eq:eg:cluster}. For $`g\leq H(\RX_2)=1$, the trivial partition $\Set{\Set{0,1,2}}$ is optimal, i.e., $\hat h_{`g}(V)=h_{`g}[\Set {\Set {0,1,2}}]$. For $`g\in [H(\RX_2),H(\RX_1)]=[1,2]$, the partition $\Set {\Set {0,1},\Set {2}}$ is optimal. For $`g\geq H(\RX_1)=2$, the partition $\Set {\Set {0},\Set {1},\Set {2}}$ into singletons is optimal.

More generally, for any submodular function $h$, the set of optimal partitions to \eqref{eq:hat h} for any
$`g$ forms a lattice called the Dilworth truncation lattice, and the sequence of Dilworth truncation
lattices forms a larger lattice which is referred to as the PLP~\cite{narayanan90}.
The lattice structure is respective to the partial order on partitions, denoted as $\mcP \preceq \mcP'$, meaning that
\begin{align}
	\forall C\in \mcP, \exists C'\in \mcP'\text{ such that } C\subseteq C'.\label{eq:finer}
\end{align}
In other words, $\mcP'$ is no smaller than $\mcP$ means that $\mcP'$ is no finer than $\mcP$. We use $\prec$ to denote the strict inequality when $\mcP\neq \mcP'$. For instance, the optimal partitions in \figref{fig:eg:h} form a chain, which is a special kind of lattice:
\begin{align*}
	\Set{\Set{0,1,2}} \succ \Set {\Set {0,1},\Set {2}} \succ \Set {\Set {0},\Set {1},\Set {2}}.
\end{align*} 
The PLP turns out to be strongly polynomial-time solvable~\cite{narayanan90,nagano10}, and it resolves
the clustering problem in hand: 
\begin{Proposition}[\mbox{\cite[Corollary~3.1]{chan16cluster}}]
	\label{pro:cluster}
	For any threshold $`g\in `R$, the clusters of
	$\pzC_{`g}$~\eqref{eq:cluster} are the non-singleton elements of the finest optimal partition of \eqref{eq:hat h} with respect to the partial order \eqref{eq:finer}.
\end{Proposition}
This can be observed in \figref{fig:eg:h}. For instance, for $`g\in [H(\RX_2),H(\RX_1)]$, the partition $\Set {\Set {0,1},\Set {2}}$ is optimal and its non-singleton element $\Set {0,1}$ is a cluster (as mentioned before).

By the above proposition, the clusters can be obtained from the optimal partitions, or more precisely, the finest optimal partitions to \eqref{eq:hat h}. In general, the finest optimal partitions form a chain called the principal sequence of partitions (PSP), which is a subset of the PLP~\cite{narayanan90}. 

The PSP, however, can be a proper subset of the PLP, and the solutions in the PSP are not the only meaningful ones. In particular, the partitions in the PLP but not the PSP will be argued to enrich the solutions of the data clustering and feature selection problems. In the following, we first extend the clustering formulation of \cite{chan16cluster} to include the entire PLP as solutions:   
\begin{Definition}
	For a threshold $`g\in `R$, the extended set of clusters is defined as
	\begin{subequations}
		\label{eq:cluster:e}
	\begin{align}
			\kern-.5em \overline{\pzC}_{`g}(\RZ_V) &:= \{B\subseteq V\mid \abs {B}>1, I(\RZ_B)\geq `g, \label{eq:cluster:e:1}\\
			&\kern1em \not\exists B'\subseteq V, `0\neq \underbrace{B\cap B'\neq B'}_{\text{or equiv.\ }B\nsupseteq B'}, I(\RZ_{B'})>`g \},\kern-.5em\label{eq:cluster:e:2}
	\end{align}
	\end{subequations}
	where $I(\RZ_B)$ is as defined in \eqref{eq:MMI}.
\end{Definition}
The following result shows that the extended set of clusters maps to the entire PLP as desired.
\begin{Theorem}
	\label{thm:cluster:e}
	The clusters in $\overline{\pzC}_{`g}(\RZ_V)$ are the non-singleton elements of the optimal partition of \eqref{eq:hat h}.
\end{Theorem}
\begin{Proof}
	See Appendix~\ref{sec:cluster:e}.
\end{Proof}
The difference between the two formulations is the non-existence condition in \eqref{eq:cluster:e:2}, which can be viewed as a relaxation of the inclusion-wise maximality constraint in \eqref{eq:cluster:2}.
More precisely, with Proposition~\ref{pro:cluster}, it follows that $\pzC_{`g}(\RZ_V)\subseteq \overline{\pzC}_{`g}(\RZ_V)$. However, the extended set of clusters may be strictly larger: The non-existence condition~\eqref{eq:cluster:e:2} forbids a set $B'$ with at least one element in $B$ and one element outside $B$ (i.e, $B$ bisects $B'$) while having a mutual information strictly larger than $`g$; it potentially allows $C_1, C_2\in \overline{\pzC}_{`g}(\RZ_V)$ such that $I(\RZ_{C_1})=I(\RZ_{C_2})=`g$ but $C_2\supsetneq C_1$. This allowed scenario is excluded in \eqref{eq:cluster} even if the threshold constraint is changed to non-strict inequality.\footnote{A non-strict inequality for \eqref{eq:cluster} will only shift the clustering solution very slightly, i.e., $\pzC_{`g}(\RZ_V)$ will be changed to the one-sided limit $\lim_{`g'\uparrow 0} \pzC_{`g'}(\RZ_V)$.} For instance, consider the example~\eqref{eq:eg:cluster} with \eqref{eq:eg:regress:1} and
\begin{align}
	H(\RX_1)=H(\RX_2)=1 \label{eq:eg:regress:3}
\end{align}
instead of \eqref{eq:eg:regress:2}. Then, as $`g$ increases to $1$, the set $\pzC_{`g}(\RZ_V)$ of clusters changes from $\Set{\Set{0,1,2}}$ to the empty set $`0$, i.e., we have $\pzC_1(\RZ_{\Set{0,1,2}})=`0$, which can be seen by noting that for $`g=1$, the finest optimal partition is the partition into singletons. In contrast, one can show that as $`g$ increases to $1$, the extended set $\overline{\pzC}_{1}(\RZ_{\Set {0,1,2}})$ of clusters changes from $\Set{\Set{0,1,2}}$ to further include the sets  $\Set{0,1}$ and $\Set{0,2}$, i.e., we have $\overline{\pzC}_1(\RZ_{\Set{0,1,2}})=\Set{\Set{0,1,2},\Set{0,1},\Set{0,2}}$. 
More generally, the additional clusters in the extended set can be characterized as follows:
\begin{Corollary}
	\label{cor:cluster:e}
	For any $`g\in `R$, we have $B\in\overline{\pzC}_{`g}(\RZ_V)`/\pzC_{`g}(\RZ_V)$ iff $\abs {B}>1$, $I(\RZ_B)=`g$ and
	\begin{align}
		B\cap B'=`0 \kern1em \text { or } \kern1em B'\subseteq B,\label{eq:consistent:e}
	\end{align}
	for all $B'\in \pzC_{`g}(\RZ_V)$ (or simply with $I(\RZ_{B'})>`g$).
\end{Corollary}
\begin{Proof}
	See Appendix~\ref{sec:cluster:e}.
\end{Proof}
\eqref{eq:consistent:e} means that a cluster $B\in\overline{\pzC}_{`g}(\RZ_V)`/\pzC_{`g}(\RZ_V)$ is consistent with the clusters in $\pzC_{`g}(\RZ_V)$ in the sense that such a cluster $B$ with $`g$ amount of mutual information does not break apart any cluster $B'\in \pzC_{`g}(\RZ_V)$ that has a strictly larger amount of mutual information than $`g$. 

As an illustration, the earlier example is both a Bayesian tree network and a tree PIN model with a more structured (graphical) clustering solution described in \cite[Section~IV]{chan16cluster}. Roughly speaking, the network can be regarded as the tree (chain) $1"-"0"-"2$ with equal-weight edges. The extended set of clusters returns all the subtrees, namely $0"-"1$ and $0"-"2$, as the clusters at $`g=1$, in addition to the trivial cluster consisting of all the nodes. It can be seen that the extended set of clusters give more flexibility in the sense of finding a cluster of an appropriate size for the application of interest. 

That being said, if the application of interest demands a smaller cluster than what are available at a given threshold, there is no particular reason why one should not increase the threshold to identify a cluster of the desired size. For the earlier example, even though $\Set{0,1}$ and $\Set{0,2}$ are not in the extended set of clusters for $`g<1$, they may be considered if a small cluster is desired. $\Set{1,2}$ is not preferred because it is not consistent with (breaks apart) $\Set{0,1}$, $\Set{0,2}$ and therefore $\Set{0,1,2}$, all of which have a strictly larger mutual information. 

\section{Feature Selection Formulation}
\label{sec:regress}

Let $\RX_U:=(\RX_i\mid i\in U)$  be a finite vector of mutually independent random variables
referred to as the features, and $\RY$ be a
random variable that depends on $\RX_U$. The joint distribution of $\RX_U$ and $\RY$ can be written as
\begin{align}
	P_{\RX_U \RY}=P_{\RY|\RX_U} \prod_{i\in U} P_{\RX_i}.\label{eq:indep}
\end{align}
For a non-negative integer $k$,
if we are to select $k$ features as the most relevant ones to $\RY$, then it is natural to choose the set that maximizes the mutual information
\begin{align}
	\max \Set{I(\RY\wedge \RX_B)\mid B\subseteq U, \abs {B}=k}.\label{eq:regress:1}
\end{align}
Such information-theoretic formulation for feature selection first appeared in \cite[FR$n$--$k$]{battiti94}, and will be referred to as the size-constrained formulation (since the size of the set of features to be selected is fixed). Note that
\begin{align*}
	I(\RY\wedge \RX_B) &= H(\RX_B) - H(\RX_B|\RY)\\
	&= \sum_{i\in B} H(\RX_i) - H(\RX_B|\RY)
\end{align*}
by \eqref{eq:indep}, which is supermodular in $B$ because $H(\RX_B|\RY)$ is submodular and $\sum_{i\in B} H(\RX_i)$ is modular in $B$. 

Unfortunately, maximizing a supermodular function as in~\eqref{eq:regress:1} (or minimizing a
submodular function) under a cardinality constraint is NP-hard in general as it generalizes~\cite[Section~10.4.4]{narayanan97}\cite{nagano11} the dense $k$-subgraph problem, e.g., see \cite{feige01}.
Therefore, we consider a relaxation that can be solved in strongly polynomial time: Given a
threshold $`g\in `R$, the preferred sets of features achieve the objective
\begin{align}
	f^*(`g):=\max_{B\subseteq U} \underbrace{\overbrace{I(\RY \wedge \RX_B)}^{f(B):=} - `g \abs {B}}_{f_{`g}(B):=}.\label{eq:regress}
\end{align}
Intuitively, for $`g>0$, the second term $-`g \abs {B}$ is a penalty in favor of a smaller set of
features. The closely related expression $f^*(`g)+`gk$ is the well-known Lagrangian dual of \eqref{eq:regress:1}, which can serve as an upper bound of \eqref{eq:regress:1}. 
The optimal solutions of \eqref{eq:regress:1} is related to those of the Lagrangian dual (and therefore \eqref{eq:regress}) as follows:
\begin{Proposition}
	\label{pro:regress}
	If $B^*$ is optimal to \eqref{eq:regress} for some $`g$, then it is also optimal to \eqref{eq:regress:1} with $k=\abs {B^*}$. (This holds even for dependent features, i.e., without the independence assumption~\eqref{eq:indep}.)
\end{Proposition}

\begin{Proof}
	Suppose to the contrary that there exists $B'$ with $\abs {B'}=\abs {B^*}$ but $I(\RX_{B'}\wedge
	\RY)>I(\RX_{B^*}\wedge \RY)$, then $f_{`g}(B')>f_{`g}(B^*)$, contradicting the optimality of
	$B^*$.
\end{Proof}

As an illustration, \figref{fig:eg:f} is a plot of $f^*(`g)$ against $`g$ for the example in \eqref{eq:eg:regress}. For
$`g\leq H(\RX_2)$, the entire set $\Set{1,2}$ of features is the optimal solution to
\eqref{eq:regress} achieving the maximum value of $f_{`g}(\Set{1,2})$. It is also the optimal
solution to \eqref{eq:regress:1} for $k=2$. For $`g\in [H(\RX_2),H(\RX_1)]$, the set $\Set {1}$ is
optimal to \eqref{eq:regress} and it is also the optimal solution to \eqref{eq:regress:1} for $k=1$.
For $`g\geq H(\RX_1)$, the empty set $`0$ is optimal to \eqref{eq:regress} and trivially optimal to
\eqref{eq:regress:1} for $k=0$.

The reason we regard \eqref{eq:regress} as a relaxation of \eqref{eq:regress:1} because the converse of
Proposition~\ref{pro:regress} does not hold in general, i.e., it is possible to find an
example where an optimal solution to \eqref{eq:regress:1} for some integer $k$ is not
optimal to \eqref{eq:regress} for any $`g$. Such an example is given in Appendix~\ref{sec:cg:regress}.

For a general supermodular function $f$, the set of optimal solutions to \eqref{eq:regress} for different
values of $`g$ forms a finite distributive lattice with respect to set inclusion~\cite{fujishige05}. By Birkhoff's representation theorem, the lattice can be characterized using a partial order over the elements of a partition of $V$. This structure was shown to be polynomial-time solvable structure and is called the principal partition
(PP). For the detailed definition and historical development of the concept, we refer the readers to~\cite{fujishige80,fujishige05,fujishige-pp-revisited}.\footnote{In the literature, the term PP is used to refer to both the distributive lattice and the induced (equivalent) structure consisting of a partial order defined over a partition of the ground set (hence the term PP). In this work, we follow this convention to use the term PP to refer to the distributive lattice.} In particular, the optimal solutions in
\figref{fig:eg:f} form a chain, which is a special kind of lattice:
\begin{align*}
	\Set{0,1,2} \supseteq \Set{0,1} \supseteq `0.
\end{align*}

There is a closely related relaxation in~\cite[(2)]{nagano11} of the general size-constrained submodular
function minimization problem.\footnote{The idea of the relaxation has appeared in \cite[Section~10.4.4]{narayanan97}, but instead of
	the size-constrained optimization problem~\eqref{eq:regress:1}, a closely-related density
	problem was considered.} Our relaxation~\eqref{eq:regress} is simpler. It appeared as
an  intermediate step~\cite[(3)]{nagano11} that contains all the solutions of
\cite[(2)]{nagano11} (with the non-negative submodular function therein chosen to be $B\mapsto H(\RX_B|\RY)$).
Another difference is that we consider the entire PP as solutions to the feature selection problem while \cite{nagano11} restricts only to the inclusion-wise maximal and minimal subsets to the general size-constrained optimization. As a result, our formulation can give more optimal solutions to \eqref{eq:regress:1} that are also meaningful. For example, consider \eqref{eq:eg:regress} but with \eqref{eq:eg:regress:2} replaced by \eqref{eq:eg:regress:3} $H(\RX_1)=H(\RX_2)=1$. In this case, the features $\RX_1$ and $\RX_2$ are equally good as each of them contains the same amount (1~bit) of
mutual information with $\RY$. It can be shown that, for $`g=1$, both $\Set{1}$ and $\Set{2}$ are optimal solutions to \eqref{eq:regress} (in addition to the optimal solutions $`0$ and $\Set{1,2}$). Thus, for $k=1$, both $\Set{1}$ and $\Set{2}$ are solutions to \eqref{eq:regress:1} as desired. However, the relaxation in \cite{nagano11} considers only the minimal solution $`0$ and maximal solution $\Set{1,2}$ to \eqref{eq:regress}, which therefore fails to give any solution to \eqref{eq:regress:1} for $k=1$.


\section{The Duality}
\label{sec:duality}

The solutions to the data clustering and feature selection problems can be related by the following mapping:
\begin{align}
	V=\Set{0}\cup U \kern1em\text{and}\kern1em
	\RZ_i=\begin{cases}
		\RY & i=0\\
		\RX_i & i\in U,
	\end{cases}
	\label{eq:map}
\end{align}
where $\RX_U$ satisfies \eqref{eq:indep}, and we assume $0\not\in U$ without loss of generality.

\begin{Theorem}
	\label{thm:duality}
	Under the mapping \eqref{eq:map}, we have for all $`g\in `R$ and $B\subseteq U$ that $B$ is an optimal solution to \eqref{eq:regress} iff $\Set {0} \cup B$ is an element of an optimal partition to \eqref{eq:hat h}.
\end{Theorem}
In other words, \emph{the dependent variable $\RZ_0=\RY$ is clustered with the set $B$ of selected features $\RZ_B=\RX_B$.} The duality can be observed from \figref{fig:eg} for the example in~\eqref{eq:eg:regress} using the mapping \eqref{eq:eg:cluster}, which agrees with \eqref{eq:map}. 
For $`g\leq H(\RX_2)$, the set $\Set{1,2}$ is optimal in \figref{fig:eg:f}, and its union
$\Set{0}\cup \Set{1,2}$ with $\Set{0}$ is contained by the optimal partition $\Set{\Set{0,1,2}}$ in
\figref{fig:eg:h}. For $`g\in [H(\RX_2),H(\RX_1)]$, the optimal subset $\Set{1}$ in
\figref{fig:eg:f} union $\Set{0}$ is contained by the optimal partition $\Set {\Set {0,1},\Set {2}}$
in \figref{fig:eg:h}. Finally, for $`g\geq H(\RX_1)$, the optimal partition $\Set {\Set {0},\Set
	{1},\Set {2}}$ in \figref{fig:eg:h} contains $\Set{0}\cup `0$, which is trivially the union of $\Set{0}$ and
the optimal subset $`0$ in \figref{fig:eg:f}.

As another example, consider \eqref{eq:eg:cluster} again but with \eqref{eq:eg:regress:1} and \eqref{eq:eg:regress:3}, i.e., the case when 
both features $\RX_1$ and $\RX_2$ are equally good. For $`g=1$, every subset of $\Set{1,2}$ is optimal to
\eqref{eq:regress}. In particular, the solutions $\Set{1}$ and $\Set{2}$  correspond to the partitions $\Set{\Set {0,1},\Set {2}}$ and $\Set{\Set {1},\Set
	{0,2}}$, which are optimal to \eqref{eq:hat h}. This is in alignment with Theorem~\ref{thm:duality}. Note that neither
of these
optimal partitions is the finest optimal partition, i.e., the partition $\Set{\Set{0},\Set{1},\Set{2}}$, and so Proposition~\ref{pro:cluster} dictates that neither $\Set {0,1}$ nor $\Set{0,2}$ is a cluster according to
\eqref{eq:cluster}.
Nevertheless, the duality result here is more general and the discrepancy is resolved via the extended clustering formulation in~\eqref{eq:cluster:e}, where as mentioned earlier, the sets $\Set{0,1}$ and $\Set{0,2}$ are indeed in the collection of extended clustering solutions.

Before proving the theorem, we first specialize the clustering solution under the current mapping~\eqref{eq:map} by exploiting the independence among the features~\eqref{eq:indep}. For $C\subseteq V$, define the $C$-block partition of $V$ as
\begin{align}
	\mcP_C:= \Set{C}\cup \Set{\Set {i}\mid i\in V`/C}.\label{eq:PC}
\end{align} 
\begin{Proposition}
	\label{pro:duality}
	For $`g>0$, any optimal $\mcP$ to \eqref{eq:hat h} under \eqref{eq:map} must satisfy $\mcP=\mcP_{\Set {0}\cup B}$~\eqref{eq:PC} for some $B\subseteq U$.
\end{Proposition}

\begin{Proof}
	Suppose to the contrary that an optimal $\mcP$ to \eqref{eq:hat h} contains 
	\begin{align*}
		C'\in \mcP: 0\not\in C',\abs {C'}>1.
	\end{align*}
	Define another partition of $V$ as
	\begin{align*}
		\mcP'=(\mcP`/C')\cup \Set {\Set {i}\mid i\in C'}.
	\end{align*}
	Then, the difference $h_{`g}[\mcP'] - h_{`g}[\mcP]$ is
	\begin{align*}
		\underbrace{\sum_{i\in C'} H(\RZ_i)-H(\RZ_{C'})}_{\mathclap{=0 \text { by \eqref{eq:map} and \eqref{eq:indep}.}}} - (\underbrace{\abs {C'}-1}_{>0})\underbrace{`g}_{>0} < 0,
	\end{align*}
	which contradicts the optimality of $\mcP$.
\end{Proof}

\begin{Proof}[Theorem~\ref{thm:duality}]
	we will break down the proof into three cases:
	\begin{enumerate}
		\item $`g>0$: In this case, we relate \eqref{eq:regress} and \eqref{eq:hat h} directly by
		rewriting the terms in \eqref{eq:regress} using $\mcP_{\Set {0}\cup B}$~\eqref{eq:PC} for
		$B\subseteq U$.
		\begin{align*}
			\abs {B} 
			&= \abs {U} - \abs {U`/B}\\
			&= \abs {U} - \abs {\mcP_{\Set {0}\cup B}} + 1\\
			I(Y\wedge \RX_B) &= H(\RY) +\kern1em \underbrace{H(\RX_B)}_{\mathclap{\sum_{i\in B} H(\RX_i) \text{ by \eqref{eq:indep}.}}} \kern1em - \kern1em H(\underbrace{\RY,\RX_B}_{\mathclap{\RZ_{\Set {0}\cup B} \text{ by \eqref{eq:map}.}} })\\
			&= H(\RY) + \sum_{i\in U} H(\RX_i) - \sum_{C\in \mcP_{\Set {0}\cup B}} h(C).
		\end{align*}
		Altogether, we have
		\begin{align*}
			I(\RY\wedge \RX_B)-`g\abs {B}
			= t- h_{`g}[\mcP_{\Set {0}\cup B}]
		\end{align*}
		where $t:=H(\RY)+\sum_{i\in U} H(\RX_i) -(\abs U+1)`g$. Since $t$ is independent of $B$,
		\begin{align*}
			\max_{B\subseteq U} I(\RY\wedge \RX_B)-`g\abs {B}
			= t- \min_{B\subseteq U}  h_{`g}[\mcP_{\Set {0}\cup B}].
		\end{align*}
		Since $\gamma>0$, by Proposition~\ref{pro:duality} the minimization on the R.H.S.\ above is the same as \eqref{eq:hat h}, which completes the proof of this case.
		
		\item $`g<0$: 
		By the submodularity of entropy~\eqref{eq:submodular}, we have for any disjoint $C_1,C_2\subseteq V$ that
		\begin{align*}
			h_{`g}(C_1\cup C_2) &\leq h_{`g}(C_1)+h_{`g}(C_2)-\smash{\underbrace{h_{`g}(`0)}_{\kern2em=-`g}}\\
			&\leq h_{`g}(C_1)+h_{`g}(C_2)
		\end{align*}
		for $`g\leq 0$. This implies that the trivial partition $\mcP=\Set{V}$ is an optimal partition to \eqref{eq:hat h} for $`g\leq 0$ because further partitioning $V$ will not decrease the sum in \eqref{eq:hat h}. In the current case $`g<0$, the above inequality is strict, and so further partitioning $V$ will increase the sum, and so the trivial partition is indeed the unique optimal solution.
		
		Now, $B=U$ is an optimal solution to \eqref{eq:regress} for $`g\leq 0$ because $I(\RY\wedge \RX_B)$ is non-decreasing in $B$. In the current case $`g<0$ with strict inequality, the solution is also unique because $\abs {B}$ is strictly increasing in $B$.
		Hence, under the mapping~\eqref{eq:map}, we have the desired conclusion for the current case that $V=\Set{0}\cup U$ is contained by the unique optimal partition $\mcP=\Set {V}$ of \eqref{eq:hat h} while $B=U$ is the unique optimal solution to \eqref{eq:regress}.
		
		\item $`g=0$: Suppose $B$ is optimal to \eqref{eq:regress}. Since $U$ is also optimal, we have
		\begin{align}
			I(\RY\wedge \RX_B) = I(\RY \wedge \RX_U) \label{eq:1}
		\end{align}
		which means that $(\RY,\RX_B)$ is independent of $\RX_{U`/B}$, or equivalently, by \eqref{eq:map},
		\begin{align}
			h(V)=h(\Set {0}\cup B) + h(U`/B).\label{eq:2}
		\end{align} 
		This implies that $\mcP=\Set {\Set {0}\cup B, U`/B}$ is also optimal for $`g=0$ because $h_{0}=h$ and $\mcP=\Set{V}$ is an optimal solution to \eqref{eq:hat h} as argued in the previous case.
		
		Conversely, suppose $\Set {0}\cup B$ is contained in an optimal partition $\mcP$ of \eqref{eq:hat h} for $`g=0$. Since the trivial partition $\Set {V}$ is also optimal as argued in the previous case, we have
		\begin{align*}
			h(V)=\sum_{C\in \mcP} h(C) &= h(\Set {0}\cup B) +  \sum_{C\in \mcP: 0\not\in C} h(C),
		\end{align*}
		which implies~\eqref{eq:2} that $(\RY,\RX_B)$ is independent of $\RX_{U`/B}$, or equivalently~\eqref{eq:1}. This completes the proof of the current case because $B=U$ is an optimal solution to \eqref{eq:regress} for $`g=0$ as argued in the previous case.
	\end{enumerate}	
\end{Proof}

The above proof of the duality result can be extended to a more general submodular function instead of the entropy function. Indeed, the proof of the important case $`g>0$ does not even use submodularity. Nevertheless, the independence assumption in~\eqref{eq:indep} is essential in the proof. An example is given in Appendix~\ref{sec:indep} to show that the duality can fail without the independence assumption.

\section{Conclusion}

In this work, we derived in a rigorous information-theoretic sense an intuitive duality between
data clustering and feature selection.
The intuition was that features that are clustered with the dependent variable are its most relevant
features.
We started by considering the info-clustering formulation
in \cite{chan16cluster} using the MMI proposed in \cite{chan15mi}, then extended the formulation to
give a more complete clustering solution that maps to the entire PLP. We also formulated the feature
selection problem as a size-constrained submodular function optimization and relaxed it to a form
solvable in polynomial-time by computing the PP. The general duality between the PLP and PP was
derived, giving the desired duality between data clustering and feature selection.

In the feature selection formulation, the cardinality of a set of feature was considered as the model
complexity of selecting that set of feature. However, it may be desirable to consider other cost
functions, e.g., the entropy, which reflects the actual amount of information in the set of feature.
The features may also be correlated in practice. It is an interesting, but appears non-trivial, task to extend
the current result to incorporate other cost functions for the model complexity and allow
statistical dependency among the features. 

\section*{Acknowledgments}

The authors would like to thank their colleagues at the Institute of Network Coding (INC) for their insightful comments and discussions. 

\appendices
\makeatletter
\@addtoreset{equation}{section}
\renewcommand{\theequation}{\thesection.\arabic{equation}}
\@addtoreset{Theorem}{section}
\renewcommand{\theTheorem}{\thesection.\arabic{Theorem}}
\@addtoreset{Lemma}{section}
\renewcommand{\theLemma}{\thesection.\arabic{Lemma}}
\@addtoreset{Corollary}{section}
\renewcommand{\theCorollary}{\thesection.\arabic{Corollary}}
\@addtoreset{Example}{section}
\renewcommand{\theExample}{\thesection.\arabic{Example}}
\@addtoreset{Remark}{section}
\renewcommand{\theRemark}{\thesection.\arabic{Remark}}
\@addtoreset{Proposition}{section}
\renewcommand{\theProposition}{\thesection.\arabic{Proposition}}
\@addtoreset{Definition}{section}
\renewcommand{\theDefinition}{\thesection.\arabic{Definition}}
\@addtoreset{Subclaim}{Theorem}
\renewcommand{\theSubclaim}{\theTheorem\Alph{Subclaim}}
\makeatother

\section{Proof of Theorem~\ref{thm:cluster:e} and Corollary~\ref{cor:cluster:e}}
\label{sec:cluster:e}

To prove Theorem~\ref{thm:cluster:e}, we will make use of the following property of property of the PLP:
\begin{Proposition}[\mbox{\cite{narayanan90}}]
	\label{pro:PLP}
	For $\mcP_1,\mcP_2\in \Pi(V)$ such that $`g_1<`g_2$ and $h_{`g_i}[\mcP_i]=\hat{h}_{`g_i}(V)$ for $i\in \Set{1,2}$, we have the partial order $\mcP_1\succeq \mcP_2$ defined in \eqref{eq:finer}.
\end{Proposition}
This follows from the more elaborate structure of the PLP described in \cite[Proposition~3.2 and 3.3]{chan16cluster}, which in turn follows from \cite[Theorem~3.5 and 3.7]{narayanan90}. The result is proved using the submodularity of entropy; the rest of the proof of Theorem~\ref{thm:cluster:e} will not rely on the submodularity.


We first show that any element $B\in \overline {\pzC}_{`g}$ is a non-singleton element of some optimal partition for \eqref{eq:hat h}.
\begin{itemize}
	\item  Suppose $I(\RZ_B)>`g$. It can be seen that the non-existence condition in \eqref{eq:cluster:e:2} implies the non-existence condition in \eqref{eq:cluster:2}, and so we have $B\in \pzC_{`g}(\RZ_V)$. By Proposition~\ref{pro:cluster}, $B$ is contained by the finest optimal partition for \eqref{eq:hat h} as desired.
	\item Suppose $I(\RZ_B)=`g$ instead. Let $\mcP^*$ be the finest optimal partition to \eqref{eq:hat h}. Then, for all $C\in \mcP^*:\abs {C}>1$, we have $I(\RZ_{C})>`g$ by Proposition~\ref{pro:cluster}, and so, by the non-existence condition in \eqref{eq:cluster:e:2}, we have
	\begin{align*}
		C\subseteq B \text { for all }C\in \mcP^*:B\cap C\neq `0.
	\end{align*}
	(n.b., the above holds trivially for $\abs {C}=1$.) Let
	\begin{align*}
		\mcP''&:=\Set {C\in \mcP^*\mid B\cap C\neq `0}\\
		\mcP&:=(\mcP^*`/\mcP'')\cup \Set {B}.
	\end{align*}
	It follows that $\mcP''\in \Pi(B)$ with $\abs {\mcP''}>1$ and $\mcP\in \Pi(V)$. By \eqref{eq:MMI},
	\begin{align}
		`g=I(\RZ_B) &\leq \frac{\sum_{C\in \mcP''} H(\RZ_C) - H(\RZ_B)}{\abs{\mcP''}-1}, \label{eq:t:1}
	\end{align} 
	which implies that
	\begin{align*}
		0&\leq h_{`g}[\mcP'']-h_{`g}(B) \\
		&= h_{`g}[\mcP^*]-h_{`g}[\mcP].
	\end{align*}
	It follows that $\mcP$ is also optimal to \eqref{eq:hat h} since $\mcP^*$ is optimal. This completes the proof as $B\in \mcP$ by construction.
\end{itemize}

We now show that any non-singleton element $B$ in any optimal partition $\mcP$ to \eqref{eq:hat h} is a cluster in $\overline {\pzC}_{`g}(\RZ_V)$. 
\begin{itemize}
	\item We first argue that $I(\RZ_B)\geq `g$ as required in \eqref{eq:cluster:e}. Suppose to the contrary that $I(\RZ_B)<`g$. Then, by \eqref{eq:MMI}, there exists $\mcP''\in \Pi(B):\abs {\mcP''}>1$ that satisfies
	\begin{align*}
		`g > I(\RZ_B) &= \frac{\sum_{C\in \mcP''} H(\RZ_C) - H(\RZ_B)}{\abs{\mcP''}-1}.
	\end{align*}
	Let $\mcP^*:=\mcP`/\Set{B} \cup \mcP''\in \Pi(V)$. The above inequality implies that
	\begin{align*}
		0&> h_{`g}[\mcP'']-h_{`g}(B) \\
		&= h_{`g}[\mcP^*]-h_{`g}[\mcP],
	\end{align*}
	which contradicts the optimality of $\mcP$.
	\item It remains to prove the non-existence condition in \eqref{eq:cluster:e:2}. Suppose to the contrary that $B'\subseteq V$ exists with $`0\neq B\cap B'\neq B'$ and $I(\RZ_{B'})>`g$. In particular, choose an inclusion-wise maximal $B'$, and any $`g'$ from the open interval $\in (`g,I(\RZ_{B'}))$ (which is non-empty by assumption). We have $B'\in \pzC_{`g'}(\RZ_V)$ by \eqref{eq:cluster} and the maximality of $B'$. By Proposition~\ref{pro:cluster}, $B'$ is contained by some (finest) optimal partition, say $\mcP'$, to \eqref{eq:hat h}. We will argue that $\mcP\not\succeq \mcP'$ (see \eqref{eq:finer}), which contradicts the property of the PLP in Proposition~\ref{pro:PLP} as desired. In particular, $B'\in \mcP'$ is not contained by $B$ because $B\cap B'\neq B'$ by assumption. $B'$ is not contained by any $C\in \mcP`/\Set{B}$ because $B\cap B'$ is non-empty by assumption, and $C$ does not intersect with $B$ and therefore does not contain $B\cap B'$. 
\end{itemize}

	Next, we will prove Corollary~\ref{cor:cluster:e}.
	To prove the ``only if" case, consider any $B\in \overline{\pzC}_{`g}(\RZ_V)`/\pzC_{`g}(\RZ_V)$. By definition~\eqref{eq:cluster:e}, $\abs {B}>1$ and $I(\RZ_{B})\geq `g$. If in the contrary that $I(\RZ_B)\neq`g$, i.e., $I(\RZ_B)>`g$, \eqref{eq:cluster:e} would imply \eqref{eq:cluster}, contradicting $B\not\in\pzC_{`g}(\RZ_V)$. For any $B'\in \pzC_{`g}(\RZ_V)$, we have $I(\RZ_{B'})>`g$ and so the non-existence condition in \eqref{eq:cluster:e:2} implies \eqref{eq:consistent:e} as desired.
	
	To prove the ``if" case, consider any $B$ satisfying the premise and the finest optimal partition $\mcP'$ to \eqref{eq:hat h}. Let
	\begin{align*}
		\mcP'':=\Set{C\in \mcP'\mid B\cap C\neq `0}.
	\end{align*}
	$\mcP''\in \Pi(B)$ because, 
	by Proposition~\ref{pro:cluster}, the non-singleton elements in $\mcP''$ are clusters in $\pzC_{`g}(\RZ_V)$, and so they are subsets of $B$ by \eqref{eq:consistent:e}. Thus, $\mcP:=\mcP'`/\mcP''\cup \Set{ B}$ is in $\Pi(V)$ and
	\begin{align*}
		h_{`g}[\mcP']-h_{`g}[\mcP]
		&=h_{`g}[\mcP'']-h_{`g}(B) \utag{*}\geq 0,
	\end{align*}
	which will complete the proof as this implies that $\mcP$ is an optimal partition of \eqref{eq:hat h} containing $B$, and so $B\in \pzC_{`g}(\RZ_V)$ by Theorem~\ref{thm:cluster:e}. ($B\not\in\pzC_{`g}(\RZ_V)$ because $I(\RZ_B)=`g$ as argued before.) To explain the last inequality~\uref{*}, consider the non-trivial case $\abs{\mcP''}>1$ (because, otherwise, $\mcP''=\Set{B}$ implies equality for \uref{*}). By assumption,
	\begin{align*}
		`g = I(\RZ_B) &\leq \frac{\sum_{C\in \mcP''} H(\RZ_C) - H(\RZ_B)}{\abs{\mcP''}-1},
	\end{align*}
	where the last inequality is because $\mcP''$ is a feasible solution to \eqref{eq:MMI}. Rearranging the terms give \uref{*} as desired.

\section{Counter-example for the converse of Proposition~\ref{pro:regress}}
\label{sec:cg:regress}

\begin{figure}
	\centering
	{\def\u{5.5}
		\tikzstyle{tp}=[point,red,thick]
		\tikzstyle{point}=[draw,circle,minimum size=.2em,inner sep=0, outer sep=.2em]
		\begin{tikzpicture}[x=.8em,y=.8em,>=latex]
		\draw[->] (0,0*\u) -- (0,2.5*\u) node [label=right:$f^*(`g)$] {};
		\draw[->] (0,0) -- (1.5*\u,0) node [label=right:$`g$] {};
		\foreach \i/\ya/\xa/\yb/\xb/\lp/\lb in {
			1/0/0/0/1.3/left/{$f_{`g}(\overbrace{`0}^{\pzB_0})=0$}, 
			2/1/0/0/1/left/{$f_{`g}(\overbrace{\Set{1}}^{\pzB_1})=1-{`g}$}, 
			4/2/0/0/0.666/left/{$f_{`g}(\overbrace{\Set{1,2,3}}^{\pzB_2})=2-3{`g}$},
			5/2.333/0/0/0.583/above left/{$f_{`g}(\overbrace{\Set{1,2,3,4}}^{\pzB_3})=2+`e-4{`g}$}}
		\draw[dashed,gray] (\xa*\u,\ya*\u)  node [inner sep=0,outer sep=0,label={[label distance=0em]\lp:{\scriptsize\lb}}] {} -- (\xb*\u,\yb*\u);
		\foreach \i/\ya/\xa/\yb/\xb/\lp/\lb in {
			3/1.333/0/0/0.666/above left/{$f_{`g}(\Set{1,4})=1+`e-2{`g}$}}
		\draw[red] (\xa*\u,\ya*\u)  node [inner sep=0,outer sep=0,label={[label distance=0em]\lp:{\scriptsize\lb}}] {} -- (\xb*\u,\yb*\u);
		\foreach \i/\x/\y/\lp/\lb in {t1/0.333/1/above right/{$(`e,1)$},
			t2/0.5/0.5/above right/{$(\tfrac12,\tfrac12)$},
			t3/1/0/above right/{$(1,0)$}}
		\path (\x*\u,\y*\u) node (\i) [tp,label={[label distance=0em]\lp:{\scriptsize\lb}}] {};
		
		\path (1*\u,0*\u) node [label=below:{\scriptsize}] {};
		\path (0.666*\u,0*\u) node [label=below:{\scriptsize}] {};
		\draw[-,thick,blue] (0,2.333*\u) to node [above right] {\scriptsize $\pzB_3$} (t1) to node [above right] {\scriptsize $\pzB_2$} (t2) to node [above right] {\scriptsize $\pzB_3$} (t3) to node [below] {\scriptsize $\pzB_0$} (1.3*\u,0);
		\end{tikzpicture}}
	\caption{The plot of $f^*(`g)$ vs $`g$ for \eqref{eq:eg2}.}
	\label{fig:eg2}
\end{figure}
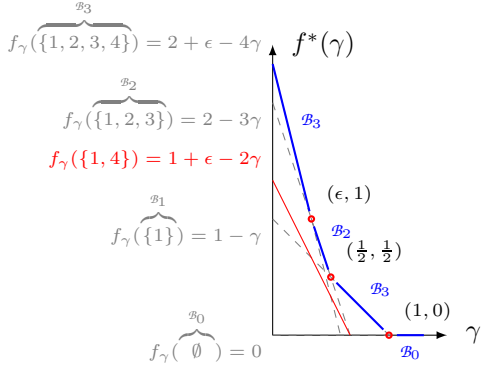


Let $U:=\Set{1,2,3,4}$ and 
\begin{align}
	\begin{split}
		\RY &:= (\RW_1,\RW_2\oplus \RW_3 \oplus \RW_4,\RW_5)\\
		\RX_1 &:= (\RW_1,\RW_2), \RX_2 := \RW_3, \RX_3 := \RW_4,
		\RX_4 := \RW_5
	\end{split}
	\label{eq:eg2}
\end{align}
where $\RW_i$'s are independent random bits with $H(\RW_i)=1$ for $i\leq 4$ and $H(\RW_5)=`e:=\frac13$, and $\oplus$ is the XOR operator. 

\figref{fig:eg2} shows the plot of $f^*(`g)$ against $`g$ and also the plots of $f_{`g}(B)$ for the following subsets $B$:
\begin{itemize}
	\item Among all the subsets $B\subseteq U$ of size $\abs{B}=1$, the choice $B=\Set{1}$ maximizes the mutual information $I(\RY\wedge \RX_B)$ to $H(\RW_1)=1$: $I(\RY\wedge \RX_i)$ is $0$ for $2\leq i\leq 3$, and it is $`e<1$ for $i=4$.
	\item Among all the subsets of size $2$, the choice $B=\Set{1,4}$ maximizes the mutual information to $H(\RW_1,\RW_5)=1+`e$: all the other mutual information are no larger than $1$.
	\item Among all the subsets of size $3$, the choice $B=\Set{1,2,3}$ maximizes the mutual information to $H(\RW_1,\RW_2+\RW_3+\RW_4)=2$: all the other mutual information are no larger than $H(\RW_1,\RW_5)=1+`e<2$.
	\item The only subset $U$ of size $4$ achieves a mutual information of $H(\RY)=2+`e$.
\end{itemize} 
It follows that $B=\Set{1,4}$ is the unique optimal solution to \eqref{eq:regress:1} for $k=2$. In \figref{fig:eg2}, it can be seen that the curve $f_{`g}(\Set{1,4})$ does not touch $f^*(`g)$, and so $B=\Set{1,4}$ is not an optimal solution to \eqref{eq:regress} for any $`g$ as desired. 

\section{Example where duality fails for dependent features}
\label{sec:indep}

Let $U:=\Set{1,2,3}$ and 
\begin{align}
	\begin{split}
		\RY &:= (\RW_1,\RW_2,\RW_3)\\
		\RX_1 &:= \RW_1, \RX_2 := (\RW_2,\RW_4), \RX_3 := (\RW_3,\RW_4)
	\end{split}
	\label{eq:eg3}
\end{align}
where $\RW_i$'s are independent random bits with $H(\RW_1)=1+`e>H(\RW_i)=1$ for $i\geq 2$ and some $`e\in (0,0.5)$. Note that the independence assumption~\ref{eq:indep} does not hold because $I(\RX_2\wedge \RX_3)=1$. 

Note that $\Set{1,2}$ and $\Set{1,3}$ are optimal solutions to \eqref{eq:regress:1} for $k=2$ but $\Set{2,3}$ is not, because
\begin{align*}
	I(\RY\wedge \RX_{\Set{1,2}})&=H(\RW_{\Set{1,2}})=2+`e\kern1em \text{and}\\
	I(\RY\wedge \RX_{\Set{1,3}})&=H(\RW_{\Set{1,3}})=2+`e\kern1em \text{but}\\
	I(\RY\wedge \RX_{\Set{2,3}})&=H(\RW_{\Set{2,3}})=2<2+`e.
\end{align*} 
By Proposition~\ref{pro:regress}, $\Set{2,3}$ cannot be optimal to \eqref{eq:regress} for any value of $`g$ either, while it can be shown that $\Set{1,2}$ and $\Set{1,3}$ are optimal solutions to \eqref{eq:regress} for $`g=1$ (in addition to the solution $\Set{1}$ and $\Set{1,2,3}$). 

Under the mapping~\eqref{eq:map}, we have
\begin{align*}
	I(\RZ_{\Set{0,1,2}})&=I(\RY,\RX_1\wedge \RX_2)=H(\RW_2)=1\\
	I(\RZ_{\Set{0,1,3}})&=I(\RY,\RX_1\wedge \RX_3)=H(\RW_3)=1\\
	I(\RZ_{\Set{0,2,3}})&=\tfrac{H(\RY)+H(\RX_2)+H(\RX_3)-H(\RY,\RX_2,\RX_3)}{2}\\ &=\tfrac{H(\RW_2)+H(\RW_3)+H(\RW_4)}2=1.5>1.
\end{align*}
It follows that neither $\Set{0}\cup \Set{1,2}$ nor $\Set{0}\cup \Set{1,3}$ is in $\overline{\pzC}_{`g}$ for any $`g\in `R$ because they fail to satisfy \eqref{eq:cluster:e:2} (with $B'=\Set{0,2,3}$) for $`g\geq 1$ and \eqref{eq:cluster:e:1} for $`g<1$. This shows that the ``only if" statement of the duality result in Theorem~\ref{thm:duality} can fail when \eqref{eq:indep} does not hold. Furthermore, it can be shown that $\Set{0}\cup \Set{2,3}$ is a cluster in $\pzC_{`g}(\RZ_V)$ (and therefore $\overline{\pzC}_{`g}(\RZ_V)$) for $`g\in [1,1.5)$. Hence, the ``if" statement of Theorem~\ref{thm:duality} also fails to hold for this example.

\bibliographystyle{IEEEtran}
\bibliography{IEEEabrv,ref}

\end{document}